\documentclass[conference, 9pt]{IEEEtran}

\usepackage{amsmath,amssymb,epsfig,multicol}
\usepackage{epsf}

\IEEEoverridecommandlockouts

\begin{document}

\title{Opportunism in Multiuser Relay Channels:\\ Scheduling, Routing and Spectrum Reuse}
\author{{\"Ozg\"ur Oyman},~\IEEEmembership{Member,~IEEE} \\ \\
\authorblockA{
Intel Research (email: ozgur.oyman@intel.com)}
}


%


\maketitle

\begin{abstract}
In order to understand the key merits of multiuser diversity techniques in relay-assisted cellular multihop networks, this paper analyzes the spectral efficiency of opportunistic (i.e., channel-aware) scheduling algorithms over a fading multiuser relay channel with $K$ users in the asymptotic regime of large (but finite) number of users. Using tools from extreme-value theory, we characterize the limiting distribution of spectral efficiency focusing on Type I convergence and utilize it in investigating the large system behavior of the multiuser relay channel as a function of the number of users and physical channel signal-to-noise ratios (SNRs). Our analysis results in very accurate formulas in the large (but finite) $K$ regime, provides insights on the potential performance enhancements from multihop routing and spectrum reuse policies in the presence of multiuser diversity gains from opportunistic scheduling and helps to identify the regimes and conditions in which relay-assisted multiuser communication provides a clear advantage over direct multiuser communication.
\end{abstract}

\section{Introduction} 

We consider the uplink and downlink of a cellular multihop/mesh system (e.g., IEEE 802.16j systems), with one base station, one fixed relay station and $K$ users. The role of the relay station is to enhance end-to-end link quality in terms of capacity, coverage and reliability \cite{Oyman_mag07}, and its presence allows the base station to choose between (i) sending/receiving data directly to/from a given user, (ii) communicating over a two-hop route where the base station sends data to the relay station and the relay station forwards the data to the users in downlink, and vice versa for the uplink. We refer to this communication model as the {\it multiuser relay channel}; which includes both the multiaccess relay channel (uplink) \cite{Kramer05, Laneman_ciss} and broadcast relay channel (downlink) \cite{Kramer05, Liang07}. 

For fixed portable applications, where radio channels are slowly varying, multiple access methods based on opportunistic scheduling mechanisms take advantage of variations in users' channel quality and allocate resources such that the user with the best channel quality is served at any given time or frequency (could be subject to certain fairness and delay constraints). It has been shown by the pioneering works \cite{Knopp95}-\nocite{Hanly98}\cite{Viswanath02} that the sum capacity under such opportunistic scheduling algorithms increases with the number of users; yielding {\it multiuser diversity} gains by exploiting the time and frequency selectivity of wireless channels as well as the independent channel variations across users. 

{\bf Contributions:} While multiuser diversity concepts over traditional cellular systems is well understood, there are open issues on the design and analysis of resource allocation and opportunistic scheduling algorithms in relay-assisted cellular multihop networks. Recently, low-complexity suboptimal resource allocation algorithms for cellular multihop networks were proposed in \cite{Oyman_asil06} based on centralized scheduling using end-to-end link quality metrics, and were shown to simultaneously realize gains from both multiuser diversity and multihop relaying to enhance capacity and coverage, provided the availability of closed-loop transmission mechanisms. 

In this paper we analyze the spectral efficiency of the opportunistic scheduling algorithms (those proposed in \cite{Oyman_asil06} as well as others) over a multiuser relay channel in the asymptotic regime of large (but finite) number of users. In this regime, merely investigating the asymptotic capacity scaling as $K \rightarrow \infty$ (which was the standard approach in works on traditional cellular systems) limits the scope of the probabilistic analysis and furthermore does not provide accurate insights on the large system behavior of the multiuser relay channel as the cellular backhaul link (e.g., link between the base station and relay station) capacity does not scale with increasing number of users. Instead, our approach involves using tools from extreme-value theory to approximate the {\it distribution} of spectral efficiency, which is tight provided that $K$ is large enough. In particular, we characterize the limiting distribution of spectral efficiency focusing on Type I convergence and utilize it in investigating the large system behavior of the multiuser relay channel as a function of the number of users and physical channel signal-to-noise ratios (SNRs). Our analysis results in very accurate formulas in the large (but finite) $K$ regime, provides insights on the role of multihop routing and spectrum reuse policies in leveraging the system-level performance of cellular systems in the presence of multiuser diversity gains from opportunistic scheduling and helps to identify the regimes and conditions in which relay-assisted multiuser communication provides a clear advantage over direct multiuser communication.


\section{Extreme-Value Theoretic Preliminaries}

Extreme-value theory deals with the stochastic behavior and asymptotic distributions of extreme values, such as maxima, minima or order statistics in general. For instance, the extreme-value distributions for maxima are obtained as the limiting distributions of greatest values in random samples of increasing size, and are used as an approximation to model the maxima of long (finite) sequences of random variables. To obtain a non-degenerate limiting distribution, it is necessary to reduce the actual greatest value by applying a linear transformation (or any transformation) with normalizing coefficients which depend on the sample size.

Let $\xi_1, \xi_2,...,\xi_M$ be independently and identically distributed (i.i.d.) random variables drawn from a common cumulative distribution function (CDF) $F(x)$ and denote the maximum of the sequence by $X_M = \max_{m=1,...,M} \xi_m$. If there exist sequences of constants $a_M > 0$, $b_M$, and some nondegenerate distribution function $\mu$ such that $(X_M - b_M)/a_M$ converges in distribution to $\mu$ as $M \rightarrow \infty$, then $\mu$ belongs to one of the three families of extreme-value distributions \cite{Leadbetter83}: Frechet, Weibull and Gumbel distributions. The distribution function of $\xi_m$, $F$, determines the exact limiting distribution. Hence, in the asymptotic regime of large $M$, it is not necessary to know the detailed nature of $F$ in order to characterize the statistical behavior of $X_M$; and it is sufficient to know which limiting distribution (if any) $F$ gives rise to, i.e. to which one of the three domains of attraction $F$ belongs. 

In \cite{vonMises36}-\cite{Gnedenko92}, von Mises (1936) and Gnedenko (1943) determine the necessary and sufficient conditions for weak convergence of $X_M$ to each of the three types of asymptotic limiting distributions, given the parent CDF $F$ corresponding to $\xi_m$. These conditions involve the tail behavior of $1-F(x)$ as $x$ increases - for each type of limit. For the case in which $F(x)$ has a density function $f(x)$; the sufficient conditions are due to von Mises \cite{vonMises36}. 

In this paper, we shall mainly be concerned with the case for which wireless fading results in received channel power distributions following the Type I extreme-value distribution, where $\mu(x)=\exp(-\exp(-x))$; and this distribution function is known as the Gumbel distribution. Assuming absolutely continuous parent distributions $F$ with density $f$, and that there exists a real number $x_1$ and $x_2 \leq \infty$ with $x_2 = \sup \{ x: F(x) < 1 \}$ such that, for all $x_1 \leq x < x_2$, $f$ has a negative derivative $f'$ and $f(x)=0$ for $x \geq x_2$, we can use the following sufficient condition to determine if the parent distribution function $F(x)$ belongs to the Type I domain of attraction (Theorem 1.6.1 in \cite{Leadbetter83}):
\begin{equation}
\lim_{x \rightarrow x_2} \frac{f'(x)(1-F(x))}{f^2(x)} = -1
\label{type1_cond}
\end{equation}
Many fading distributions, e.g. Rayleigh, Rician, lognormal, could be given as examples leading to Type I convergence. We can determine the normalizing constants $a_M$ and $b_M$ for Type I convergence, by solving for $b_M$ in $1-F(b_M)=1/M$ and setting $a_M = \eta(b_M)$, where $\eta$ is the reciprocal hazard function $\eta(x) = (1-F(x))/f(x)$. 

Defining the spectral efficiency function as $C(x)=\log_2(1+x)$, the following lemma extends the convergence results on maxima of $\{\xi_m\}_{m=1}^M$ to the maxima of spectral efficiencies by determining the constants necessary to preserve the same limiting extreme-value distribution. The simple proof (omitted) exploits the fact that $C(x)$ is a monotonically increasing function of $x$.

{\bf Lemma 1:} {\it Assume that there exist sequences of constants $a_M$, $b_M$ (the choice will depend on the distribution $F$) such that 
\begin{equation}
{\Bbb P} \left( \frac{X_M - b_M}{a_M} \leq x \right) \rightarrow \mu(x)
\label{cond_max1}
\end{equation}
as $M \rightarrow \infty$, for some limiting extreme-value distribution $\mu$. Now, let $Y_M = \max_m C(\mathsf{SNR} \, \xi_m)$ for a given constant $\mathsf{SNR}>0$. If (\ref{cond_max1}) holds, then there exist sequences of constants $c_M$, $d_M$ such that 
\begin{equation}
{\Bbb P} \left( \frac{Y_M - d_M}{c_M} \leq x \right) \rightarrow \mu(x)
\label{cond_max2}
\end{equation}
where
\begin{eqnarray}
c_M = \frac{\log_2(e)\, \mathsf{SNR}\, a_M}{1+ \mathsf{SNR} \, b_M}  & d_M = \log_2(1+\mathsf{SNR} \, b_M) \nonumber
\end{eqnarray}
provided that $c_M \rightarrow 0$ as $M \rightarrow \infty$.}

\section{Multiuser Relay Channel Model}

\begin{figure}[t]  

 \centering

  \includegraphics[height=!,width=2.6in]{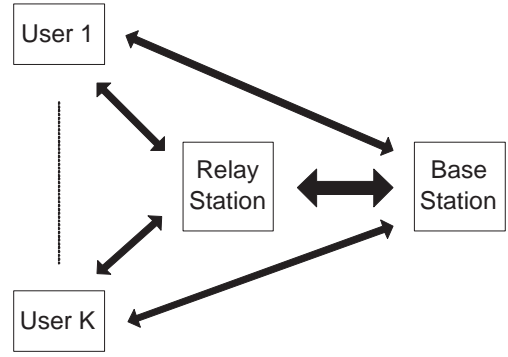}

  \caption{Multiuser relay channel model (both uplink and downlink).}

  \label{micro_net}

\end{figure}

Consider the network depicted in Fig. \ref{micro_net} with $K+2$ nodes, in which $K$ users, indexed by $k=1,...,K$, send/receive information to/from a base station. The relay station is designated to help users transmit/receive information utilizing its high capacity backhaul link to the base station. In other words, multiple users share a single relay for uplink and downlink (i.e. multiaccess and broadcast). 

We assume that all the links over the multiuser relay channel are corrupted by additive white Gaussian noise (AWGN). Furthermore, the links between the base station and users are assumed to be under frequency-flat multiplicative fading i.i.d. across users, with complex channel gains $\{h_k\}_{k=1}^K$, where $h_k \in {\Bbb C}$ is a complex random variable representing the channel between the base station and user $k$, drawn from an arbitrary continuous distribution $F_h$ with ${\Bbb E}[|h_k|^2]=1,\,\forall k$. The average received SNR between the base station and user $k$ equals $\mathsf{SNR}_k^{(b)}$. Analogously, the links between the relay station and users have average SNRs given by $\{\mathsf{SNR}_k^{(r)}\}_{k=1}^K$,\footnote{Superscript $b$ stands for "base" station and $r$ stands for "relay" station.} and are under frequency-flat fading i.i.d. across users, with complex channel gains $\{g_k\}_{k=1}^K$, where $g_k \in {\Bbb C}$ represents the channel between the relay station and user $k$, drawn from an arbitrary continuous distribution $F_g$ with ${\Bbb E}[|g_k|^2]=1,\,\forall k$. The set of channels $\{h_k\}_{k=1}^K$ and $\{g_k\}_{k=1}^K$ are independent. It is assumed that the cellular backhaul link between the base station and relay station is an AWGN line-of-sight (LOS) connection with received SNR equal to $\mathsf{SNR}_{B}$. The fading states over the multiuser links remain constant during the transmission of a codeword (slow fading assumption). Due to slow fading, each terminal in the multiuser relay channel is considered to obtain full channel state information regarding its transmission/reception links.  

For the following analysis, we assume a time-division based (half duplex) relaying constraint for multi-hop routing protocols, which is due to the practical limitation that terminals can often not transmit and receive at the same time. In particular, we consider a two-phased decode-and-forward protocol; where, for a given routing path between the base station and a given user, the relay station hears and fully decodes the transmitted data signal in the first phase and forwards its re-encoded version in the second phase.

\section{Relaying in the Presence of Multiuser Diversity}

Consider the scheduling problem such that $K$ users in the multiuser relay channel are to be assigned time-slots for transmission/reception over a common bandwidth. In particular, we are interested in the max-route centralized scheduling algorithm proposed by \cite{Oyman_asil06}, which is an extension of the maximum signal-to-interference-plus-noise ratio (max-SINR) algorithm to relay-assisted networks. Analogous to the max-SINR scheduler, which always serves the user with the best instantaneous rate at any given time/frequency resource, the max-route scheduling algorithm schedules the user with the highest end-to-end instantaneous rate. More specifically, the algorithm works as follows in the multiuser relay channel: (i) The base station compares the instantaneous rates over the direct links between itself and the users determined by $\{h_k\}_{k=1}^K$ and finds the best user. (ii) The relay station compares the instantaneous rates over the links between itself and the users determined by $\{g_k\}_{k=1}^K$ and informs the base station about the index and instantaneous rate corresponding to the best user. (iii) The base station compares the link quality of the best user over its direct link with the end-to-end link quality of the best user over the relay-assisted multihop route (accounting for the transmission over the cellular backhaul connection between the base station and relay station assuming equal time-sharing between relay-user link and wireless backhaul link), and schedules the user with the highest instantaneous rate. Here, we have dropped the dependence of average received SNRs on the user index $k$ and set $\mathsf{SNR}_k^{(b)} = \mathsf{SNR}^{(b)}$ and $\mathsf{SNR}_k^{(r)} = \mathsf{SNR}^{(r)},\,\forall k$. 

Assuming Gaussian inputs, i.e., all input signals have the temporally i.i.d. zero-mean circularly symmetric complex Gaussian distribution, the maximum supportable end-to-end spectral efficiency over the multiuser relay channel achieved by the max-route opportunistic scheduling algorithm is given by (in bits per second per Hertz (bps/Hz)) 
\begin{eqnarray}
\mathsf{C}^{\mathrm{route}} & = & \max \left[ \max_k C(\mathsf{SNR}^{(b)} |h_k|^2), \right. \nonumber \\
& & \left. \frac{1}{2} \min \left( C(\mathsf{SNR}_B), \max_k C(\mathsf{SNR}^{(r)} |g_k|^2 ) \right) \right],
\label{spec_eff_max}
\end{eqnarray}
where the $1/2$ penalty factor for the end-to-end spectral efficiency of the relay-assisted multihop route from the base station to the user is due to the half-duplex constraint imposed at the relay station. 
In (\ref{spec_eff_max}), we have two maxima among $K$ i.i.d. spectral efficiency random variables, however the maximum corresponding to the relay-user link is truncated due to the presence of the constant capacity backhaul link between the base station and relay station. The end-to-end spectral efficiency is represented as the maximum of the spectral efficiency of the direct link and that of the relay-assisted two-hop link, which implies that the users are scheduled in the presence of a channel-adaptive (i.e., dynamic) multihop routing mechanism which finds their best route to the base station (i.e. switching between direct transmission and two-hop relaying).

As the result of Lemma 1, in the asymptotic regime of large $K$, there exist sequences of constants $c_K^{(h)}$, $d_K^{(h)}$, $c_K^{(g)}$, $d_K^{(g)}$ such that the limiting distribution of $\mathsf{C}^{\mathrm{route}}$ is expressed as
\footnote{ $\, \, \, \longrightarrow^{\!\!\!\!\!\!\!\mbox{\tiny d}}\,\,\,\,$ denotes convergence in distribution.}
\begin{eqnarray}
\mathsf{C}^{\mathrm{route}} & \, \, \, \longrightarrow^{\!\!\!\!\!\!\!\mbox{\tiny d}}\,\,\,\,\, & \max \left[ c_K^{(h)} \Theta^{(h)} + d_K^{(h)}, \right. \,\,\,\,\,\,\,\,\,\,\,\,\,\,\,\,\,\,\,\,\,\,\,\,\,\,\,\,\,\,\,\,\,\,\,\,\,\,\,\,\,\,\, \nonumber \\ & & \left. 
\frac{1}{2} \min \left( C(\mathsf{SNR}_B), c_K^{(g)} \Theta^{(g)} + d_K^{(g)} \right) \right]
\label{spec_eff_conv}
\end{eqnarray}
where the independent random variables $\Theta^{(h)}$ and $\Theta^{(g)}$ follow the extreme-value distributions $\mu^{(h)}$ and $\mu^{(g)}$ (which may be the same or different), respectively, which in turn are determined by fading distributions $F_h$ and $F_g$.  

{\bf Theorem 1:} {\it In the asymptotic regime of large $K$, assuming that $F_h$ and $F_g$ both belong to Type I domain of attraction (i.e. both $\mu^{(h)}$ and $\mu^{(g)}$ follow the Gumbel distribution), the probability that the direct link yields higher spectral efficiency than relay-assisted two-hop routing is specified by $P_{{\cal C}} = \lim_{K \rightarrow \infty} {\cal P}_K$ such that
\begin{eqnarray}
P_{{\cal C}} & = & {\Bbb P} \left( {\cal C} \cap {\cal A} \right) + {\Bbb P} \left( {\cal C}| {\cal A}^c \right) (1-{\Bbb P} ({\cal A})), 
\label{thr1_prb}\\
{\cal P}_K & = & {\Bbb P} \left( \max_k C(\mathsf{SNR}^{(b)} |h_k|^2) \right. \nonumber \\ & > &
\left. \frac{1}{2} \min \left( C(\mathsf{SNR}_B), \max_k C(\mathsf{SNR}^{(r)} |g_k|^2 ) \right) \right),
\label{emp_prob}
\end{eqnarray}
where events ${\cal A}$ and ${\cal C}$ are defined as
\begin{eqnarray}
{\cal A} & = & \left\{w:\,c_K^{(g)}\Theta^{(g)}(w) + d_K^{(g)} \leq C(\mathsf{SNR}_B) \right\}, \nonumber \\
{\cal C} & = & \left\{w:\,c_K^{(h)}\Theta^{(h)}(w) + d_K^{(h)} \right. \nonumber \\ 
& > & \left. \frac{1}{2} \min \left( C(\mathsf{SNR}_B), c_K^{(g)}\Theta^{(g)}(w) + d_K^{(g)} \right) \right\}, \nonumber 
\end{eqnarray}
the corresponding probabilities are given by ${\Bbb P}({\cal A}) = e^{- z_{K,2}}$, ${\Bbb P} \left( {\cal C} | {\cal A}^c \right) = 1-e^{-z_{K,3}}$ and equation (\ref{quad1}), with sequences $z_{K,1}$, $z_{K,2}$ and $z_{K,3}$ such that
\begin{eqnarray}
z_{K,1} & = & \exp \left( \frac{d_K^{(g)}-2d_K^{(h)}}{c_K^{(g)}} \right) \nonumber \\
z_{K,2} & = & \exp \left( \frac{d_K^{(g)}-C(\mathsf{SNR}_B)}{c_K^{(g)}} \right) \nonumber \\
z_{K,3} & = & \exp \left( \frac{d_K^{(h)}-C(\mathsf{SNR}_B)/2}{c_K^{(h)}} \right), \nonumber              
\end{eqnarray}
and function $Q$ defined as $Q(x) = \frac{1}{\sqrt{2\pi}} \int_x^{\infty} e^{-y^2/2}\,dy$.} 
\begin{figure*}[!ht]
\begin{equation}
{\Bbb P} \left( {\cal C} \cap {\cal A} \right) = \sqrt{\frac{\pi}{z_{K,1}}}\, \exp \left(\frac{1}{4z_{K,1}} \right) \, Q \left( \, z_{K,3} \, \sqrt{2\,z_{K,1}} + \frac{1}{\sqrt{2z_{K,1}}}\right)  +  e^{-z_{K,2}} \left(1- e^{-z_{K,3}} \right) \label{quad1}
\end{equation}
\hrule
\end{figure*}

\begin{figure}[t]  

 \centering

  \includegraphics[height=!,width=3in]{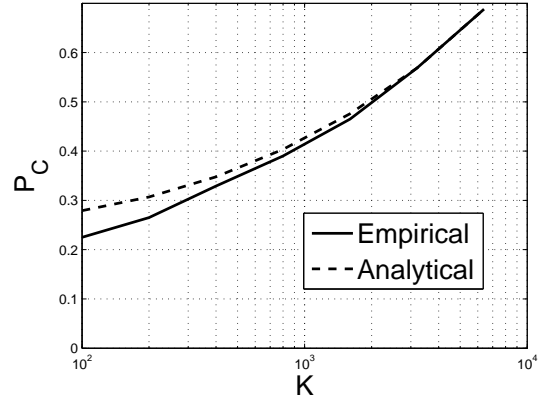}

  \caption{Probability that the direct link yields higher spectral efficiency than relay-assisted two-hop routing as a function of the number of users.}

  \label{plot1}

\end{figure}

Assuming Rayleigh fading distribution on $F_h$ and $F_g$, and setting $\mathsf{SNR}^{(b)} = 1$, $\mathsf{SNR}^{(r)} = 10$ and $\mathsf{SNR}_B = 100$, we plot in Fig. \ref{plot1} empirically generated (based on Monte Carlo simulations) probability values for (\ref{emp_prob}) as a function of the number of users $K$. We also plot our analytical result in Theorem 1 obtained by using extreme-value theory. We observe that our analytical results are well in agreement with the empirical results, and that higher level of accuracy is achieved with higher $K$. Moreover, this comparison reveals an interesting tradeoff on the merits of multihop relaying; showing that relaying becomes less beneficial as more multiuser diversity gains can be realized. In other words, the direct link spectral efficiency grows unboundedly with increasing number of users while the end-to-end spectral efficiency of the relay-assisted two-hop route converges to $C(\mathsf{SNR}_B)/2$ due to the presence of the constant capacity backhaul link, designating the use of the direct link more advantageous as $K$ grows larger. However, in many practical deployment scenarios we have $\mathsf{SNR}_B \gg \mathsf{SNR}^{(b)}$ and $\mathsf{SNR}_B \gg \mathsf{SNR}^{(r)}$ ensuring that the wireless backhaul links have high enough capacity so that they do not become the bottleneck for the multihop communication in the presence of opportunistic scheduling; enabling simultaneous realization of multiuser diversity gains along with capacity and coverage enhancements from relaying techniques particularly for cell edge users for which $\mathsf{SNR}^{(r)} \gg \mathsf{SNR}^{(b)}$.

\section{Spectrum Reuse in Multiuser Relay Channels}

\begin{figure}[t]  

 \centering

  \includegraphics[height=!,width=2.6in]{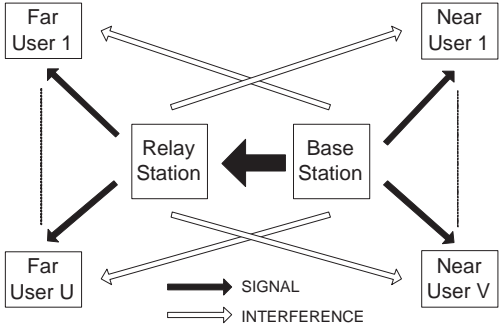}

  \caption{Multihop broadcast channel model (downlink).}

  \label{micro_net3}

\end{figure}

In this section, with cellular applications in mind, we will focus on a special mode of relay-assisted downlink communication over the multiuser relay channel depicted in Fig. \ref{micro_net3} with $U+V+2$ nodes ($K=U+V$ users). The users are divided into two categories: $U$ far users, indexed by $u=1,...,U$, with poor quality links to the base station (e.g., users at cell edge) and $V$ near users, indexed by $v=1,...,V$, with high quality links to the base station\footnote{In contrast to Section IV, this section assumes that a network entry and handoff algorithm has been executed prior to the designation of far and near users and that there is no dynamic routing mechanism for a given user to switch between direct link and relay-assisted two-hop route.}. The role of the relay station is to enhance end-to-end link quality for the far users in terms of capacity and coverage using multihop routing techniques \cite{Oyman_mag07} and its presence allows the base station to communicate with the far users over a two-hop route where the base station sends data to the relay station over a high capacity wireless backhaul link and the relay station forwards the data to the far users with possible interference from the base station over the second hop. Meanwhile, the near users receive downlink data from the base station directly (with no help from the relay station) over the same bandwidth with possible interference from the relay station. We refer to this communication model as the {\it multihop broadcast channel}; which is a special case of the more general broadcast relay channel.

While we adopt all the general assumptions in Section II on channel modeling and statistics, we now introduce a more specific setup to simplify analysis and address the distinction between far and near user models. In particular, for the links between the base station and users, we drop the dependence of average received SNRs on the user index $k$ by setting $\mathsf{SNR}_k^{(b)} = \mathsf{SNR}_F^{(b)}$ for $U$ far users and $\mathsf{SNR}_k^{(b)} = \mathsf{SNR}_N^{(b)}$ for $V$ near users, and similarly for the links between the relay station and users, we set $\mathsf{SNR}_k^{(r)} = \mathsf{SNR}_F^{(r)}$ for $U$ far users and $\mathsf{SNR}_k^{(r)} = \mathsf{SNR}_N^{(r)}$ for $V$ near users\footnote{Subscript $F$ stands for "far" and $N$ stands for "near".}. For notational clarity, we re-write the complex channel gains as follows: Channel links between the base station and far users are represented by $\{h_{F,u}\}_{u=1}^{U}$, those between the base station and near users are represented by $\{h_{N,v}\}_{v=1}^V$, those between the relay station and far users are represented by $\{g_{F,u}\}_{u=1}^U$, and those between the relay station and near users are represented by $\{g_{N,v}\}_{v=1}^V$. 

\begin{figure}[t]  

 \centering

  \includegraphics[height=!,width=2.9in]{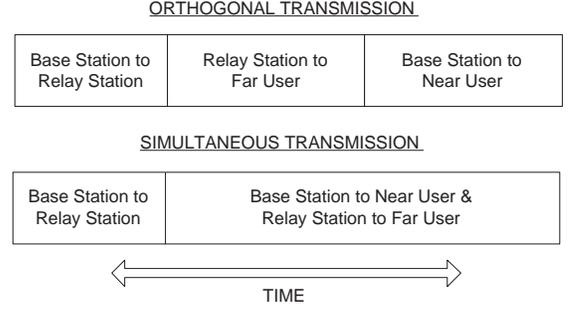}

  \caption{Multihop time allocation policies based on orthogonal and simultaneous transmissions.}

  \label{micro_net4}

\end{figure}

Consider the scheduling problem such that $U$ far users and $V$ near users are to be assigned time slots for reception over a common bandwidth. This problem involves transmissions over three types of links: (i) $L_B$: Wireless backhaul link between the base station and relay station, (ii) $L_F$: The link between the relay station and far users and (iii) $L_N$: The link between the base station and near users. We assign positive time-sharing coefficients $\beta_B$, $\beta_F$ and $\beta_N$ to links $L_B$, $L_F$ and $L_N$, respectively, such that $\beta_B + \beta_F + \beta_N = 1$ and define the following time-allocation policies as depicted in Fig. \ref{micro_net4}:

{\it a) Orthogonal transmission (no spectrum reuse):} Links $L_B$, $L_F$ and $L_N$ are active over different time resources with the corresponding time-sharing constants $\beta_B$, $\beta_F$ and $\beta_N$. The relay station compares the channel gains $\{g_{F,u}\}_{u=1}^U$ of the far users and assigns link $L_F$ for downlink transmission to the far user with the highest instantaneous rate. Analogously, the base station compares the channel gains $\{h_{N,v}\}_{v=1}^V$ of the near users and assigns link $L_N$ for downlink transmission to the near user with the highest instantaneous rate.

{\it b) Simultaneous transmission (spectrum reuse):} Link $L_B$ is active over $\beta_B$ fraction of the time, while links $L_F$ and $L_N$ are simultaneously active over $\beta_F+\beta_N$ fraction of the time. The relay station accounts for average received signal-to-noise ratios $\mathsf{SNR}_F^{(r)}$, $\mathsf{SNR}_F^{(b)}$ and channel gains $\{g_{F,u}\}_{u=1}^U$, $\{h_{F,u}\}_{u=1}^U$ to schedule link $L_F$ for downlink transmission to the far user with the highest instantaneous rate. Similarly, the base station accounts for $\mathsf{SNR}_N^{(b)}$, $\mathsf{SNR}_N^{(r)}$, $\{h_{N,v}\}_{v=1}^V$, $\{g_{N,v}\}_{v=1}^V$ to schedule link $L_N$ for downlink transmission to the near user with the highest instantaneous rate. No coordination is assumed to be present between the base station and relay station to manage the resulting intracell interference.

We now study the spectral efficiency performance of these two downlink transmission protocols over the multihop broadcast channel in the asymptotic regime of large (but finite) $U$ and $V$:

\subsection{Orthogonal Transmissions}

The maximum supportable end-to-end spectral efficiency over the multihop broadcast channel achieved by the max-SINR opportunistic scheduling algorithm in the presence of the orthogonal transmission protocol is given by (in bits per second per Hertz (bps/Hz)) 
\begin{eqnarray}
\mathsf{C}^{\mathrm{ort}} & = & \beta_N \max_{v=1,...,V} C(\mathsf{SNR}_N^{(b)} |h_{N,v}|^2 ) \nonumber \\
& + & \min \left[ \beta_B \, C(\mathsf{SNR}_B),\, \beta_F \max_{u=1,...,U} C(\mathsf{SNR}_F^{(r)} |g_{F,u}|^2 ) \right], \nonumber \\
\label{spec_eff_orth}
\end{eqnarray}
Our goal is to characterize the extreme-value distribution of $\mathsf{C}^{\mathrm{ort}}$ in the limit of large $U$ and $V$, i.e., large number of far/near users. As the result of Lemma 1, in the asymptotic regime of large $U$, $V$, there exist sequences of constants $c_U^{(g)}$, $d_U^{(g)}$, $c_V^{(h)}$, $d_V^{(h)}$ such that the limiting distribution of $\mathsf{C}^{\mathrm{ort}}$ is expressed as
\begin{eqnarray}
\mathsf{C}^{\mathrm{ort}} & \, \, \longrightarrow^{\!\!\!\!\!\!\!\mbox{\tiny d}}\,\,\,\, & \beta_N \left( c_V^{(h)} \Theta^{(h)} + d_V^{(h)} \right) \,\,\,\,\,\,\,\,\,\,\,\,\,\,\,\,\,\,\,\,\,\,\,\,\nonumber \\ & + &  
\min \left[ \beta_B \, C(\mathsf{SNR}_B), \, \beta_F \left( c_U^{(g)} \Theta^{(g)} + d_U^{(g)} \right) \right]. \nonumber
\end{eqnarray}

{\bf Theorem 2:} {\it In the asymptotic regime of large $U$, $V$, assuming that $F_h$ and $F_g$ both belong to Type I domain of attraction (i.e. both $\mu^{(h)}$ and $\mu^{(g)}$ follow the Gumbel distribution), the average spectral efficiency of the orthogonal transmission protocol defined as $\Xi^{\mathrm{ort}}={\Bbb E}\,[\mathsf{C}^{\mathrm{ort}}]$ can be expressed as 
\begin{equation}
\Xi^{\mathrm{ort}} = \beta_N \left( c_V^{(h)} \kappa + d_V^{(h)} \right) + \beta_B \, C(\mathsf{SNR}_B) -  \beta_F\,c_U^{(g)}\,\mathrm{Ei}(z_{U,1}) 
\label{cap_orth}
\end{equation}
where $\kappa \approx 0.57721566$ is Euler's constant (since ${\Bbb E}[\Theta^{(h)}]=\kappa$ under Type I convergence), $\mathrm{Ei}(x)$ is the exponential integral function defined by $\mathrm{Ei}(x)=\int_x^{\infty} \frac{e^{-y}}{y} \,dy$ and}
$$
z_{U,1}=\exp\left(\frac{\beta_F\,d_U^{(g)}-\beta_B\,C(\mathsf{SNR}_B)}{\beta_F\,c_U^{(g)}}\right).
$$ 

\subsection{Simultaneous Transmissions}

The maximum supportable end-to-end spectral efficiency over the multihop broadcast channel achieved by the max-SINR opportunistic scheduling algorithm in the presence of the simultaneous transmission protocol is given by (in bits per second per Hertz (bps/Hz)) 
\begin{eqnarray}
\mathsf{C}^{\mathrm{sim}} & = & (\beta_F + \beta_N) \max_{v=1,...,V} C(\mathsf{SINR}_{N,v}) \nonumber \\
& + & \min \left[ \beta_B \, C(\mathsf{SNR}_B), (\beta_F+\beta_N) \max_{u=1,...,U} C(\mathsf{SINR}_{F,u}) \right] \nonumber \\
\label{spec_eff_sim}
\end{eqnarray}
where $\mathsf{SINR}_{F,u}$ is the SINR for far user $u$ and $\mathsf{SINR}_{N,v}$ is the SINR for near user $v$ given by
\begin{eqnarray}
\mathsf{SINR}_{F,u} & = & \frac{\mathsf{SNR}_F^{(r)} |g_{F,u}|^2}{\mathsf{SNR}_F^{(b)}|h_{F,u}|^2 + 1}, \,\,\,\,\, u=1,...,U \nonumber \\
\mathsf{SINR}_{N,v} & = & \frac{\mathsf{SNR}_N^{(b)} |h_{N,v}|^2}{\mathsf{SNR}_N^{(r)}|g_{N,v}|^2 + 1}, \,\,\,\,\, v=1,...,V \nonumber
\end{eqnarray}
As the result of Lemma 1, in the asymptotic regime of large $U$, $V$, there exist sequences of constants $\rho_U^{(f)}$, $\sigma_U^{(f)}$, $\rho_V^{(n)}$, $\sigma_V^{(n)}$ such that the limiting distribution of $\mathsf{C}^{\mathrm{sim}}$ is expressed as
\begin{eqnarray}
\mathsf{C}^{\mathrm{sim}} & \, \, \, \longrightarrow^{\!\!\!\!\!\!\!\mbox{\tiny d}}\,\,\,\,\, & (\beta_F + \beta_N)\, \left( \rho_V^{(n)} \Theta^{(n)} + \sigma_V^{(n)} \right) \,\,\,\,\,\,\,\,\,\,\,\,\,\,\,\,\,\,\,\,\,\,\,\,\,\,\,\,\,\,\,\,\,\,\,\,\,\,\,\,\,\,\, \nonumber \\ & + &  
\min \left[ \beta_B \, C(\mathsf{SNR}_B), (\beta_F+\beta_N)\,\left(\rho_U^{(f)} \Theta^{(f)} + \sigma_U^{(f)} \right) \right] \nonumber
\end{eqnarray}
where random variables $\Theta^{(f)}$ and $\Theta^{(n)}$ follow the extreme-value distributions $\mu^{(f)}$ and $\mu^{(n)}$, respectively, resulting from the convergence of the maxima of $\{\mathsf{SINR}_{F,u}\}_{u=1}^U$ and $\{\mathsf{SINR}_{N,v}\}_{v=1}^V$.

{\bf Theorem 3:} {\it In the asymptotic regime of large $U$, $V$, assuming that both $\mu^{(f)}$ and $\mu^{(n)}$ follow the Gumbel distribution, the average spectral efficiency of the simultaneous transmission protocol defined as $\Xi^{\mathrm{sim}}={\Bbb E}\,[\mathsf{C}^{\mathrm{sim}}]$ can be expressed as 
\begin{eqnarray}
\Xi^{\mathrm{sim}} & = & (\beta_F+\beta_N) \, \left( \rho_V^{(n)}\kappa+ \sigma_V^{(n)} \right) \nonumber \\
& + & \beta_B\,C(\mathsf{SNR}_B) -(\beta_F+\beta_N) \, \rho_U^{(f)}\,\mathrm{Ei}(z_{U,2}) \label{cap_sim}
\end{eqnarray}
where} 
$$
z_{U,2}=\exp\left(\frac{(\beta_F+\beta_N)\,\sigma_U^{(f)}-\beta_B\,C(\mathsf{SNR}_B)}{(\beta_F+\beta_N)\,\rho_U^{(f)}}\right).
$$

\begin{figure}[t]  

 \centering

  \includegraphics[height=!,width=2.9in]{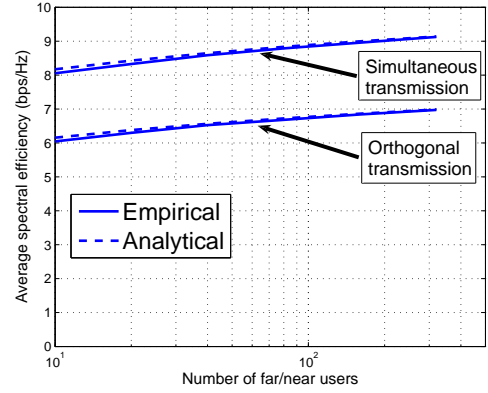}

  \caption{Average spectral efficiency as a function of the far / near users for the orthogonal and simultaneous transmission protocols.}

  \label{simul_td}

\end{figure}

Assuming Rayleigh fading distribution on $F_h$ and $F_g$ and setting $\mathsf{SNR}_B=1000$, $\mathsf{SNR}_F^{(r)}=\mathsf{SNR}_N^{(b)}=100$, $\mathsf{SNR}_F^{(b)}=\mathsf{SNR}_N^{(r)}=1$, we plot in Fig. \ref{simul_td} average spectral efficiency as a function of the number of far / near users (set $U=V,\,K=2U=2V$) for orthogonal and simultaneous transmission protocols over the multihop broadcast channel with time-sharing coefficients set to $\beta_B=0.25$, $\beta_F=0.25$, $\beta_N=0.5$. Here, we compare empirically generated average spectral efficiencies $\Xi^{\mathrm{ort}}$ and $\Xi^{\mathrm{sim}}$ with their analytical counterparts in (\ref{cap_orth}) and (\ref{cap_sim}). The empirical results are obtained by averaging the expressions in (\ref{spec_eff_orth}) and (\ref{spec_eff_sim}) over a large number of randomly generated fading realizations (based on Monte Carlo simulations). From Fig. \ref{simul_td}, we validate that our analytical results are well in agreement with the empirical results and that higher level of accuracy is achieved with higher $U,\,V$. Moreover, we observe that both orthogonal transmission and simultaneous transmission protocols realize multiuser diversity gains from opportunistic scheduling techniques due to the increase of the average spectral efficiency in the number of far / near users. Finally, we find that significant spectral efficiency gains can be achieved through spectrum reuse by the base station and relay station based on the superior performance of the simultaneous transmission protocol over the orthogonal transmission protocol. We emphasize that such spectral efficiency gains are possible despite that the simultaneous transmission protocol requires no scheduling coordination between the base station and relay station and does not enforce any policy for interference management.

\begin{footnotesize}
\renewcommand{\baselinestretch}{0.2}
\bibliographystyle{IEEE}

\end{footnotesize}


\end{document}